\documentclass[]{vgtc}
\usepackage[utf8]{inputenc}
\usepackage{titlesec}
\usepackage{parskip}
\usepackage[nolist,nohyperlinks]{acronym}
\usepackage{color}
\usepackage{pdfpages}
\usepackage{multirow}
\usepackage{dblfloatfix}
\usepackage{multicol}

\acrodef{GSR}[GSR]{Galvanic Skin Response}
\acrodef{HRV}[HRV]{Heart Rate Variability}
\acrodef{ROS}[ROS]{\textit{Robot Operating System}}
\acrodef{CGP}[CGP]{\textit{Contextual Gaussian Process}}
\acrodef{GP}[GP]{Gaussian Process}
\acrodef{GMM}[GMM]{Gaussian mixture model}
\acrodef{RBF}[RBF]{\textit{Radial Basis Function}}
\acrodef{FAA}[FAA]{Federal Aviation Administration}
\acrodef{UCB}[UCB]{\textit{upper confidence bound}}
\acrodef{RANSAC}[RANSAC]{\textit{Random Sampling and Consensus}}
\acrodef{ARD}[ARD]{\textit{Auto Relevance Determination}}
\acrodef{ADAS}[ADAS]{\textit{Advanced Driver Assistance Systems}}
\acrodef{VR}[VR]{Virtual Reality}
\acrodef{CNNs}[CNNs]{convolutional neural networks}
\acrodef{IRB}[IRB]{Institutional Review Board}
\acrodef{UCSC}[UCSC]{University of California, Santa Cruz}
\acrodef{PCA}[PCA]{Principal Component Analysis}

\acrodef{HR}[HR]{Human-Robot}
\acrodef{HSIA}[HSIA]{Human System Integration Architectures}
\acrodef{HCI}[HCIs]{human computer interactions}
\acrodef{CSE}[CSE]{cognitive state estimation}
\acrodef{CTL}[CTL]{cognitive task load}
\acrodef{CLT}[CLT]{Cognitive Load Theory}
\acrodef{TLX}[NASA-TLX]{NASA Task Load Index}
\acrodef{ML}[ML]{machine learning}
\acrodef{RL}[RL]{reinforcement learning}
\acrodef{EEG}[EEG]{Electroencephalography}
\acrodef{TRAADRE}[TRAADRE]{TRust in Autonomous ADvisors for Robotic Exploration}
\acrodef{DL}[DL]{deep learning}
\acrodef{HRI}[HRI]{human robot interaction}
\acrodef{AI}[AI]{artificially intelligent}
\acrodef{PDF}[PDF]{Probability Density Function}
\acrodef{WS}[WS]{wearable sensors}
\acrodef{CL}[CL]{cognitive load}
\acrodef{MATB-II}[MATB-II]{Multi-Attribute Task Battery II}
\acrodef{HARE}[HARE]{Human Aware Robotic Exploration}
\acrodef{MW}[MW]{Mental Workload}

\acrodef{RM}[RM]{Resource Management}
\acrodef{SM}[SM]{System Monitoring}
\acrodef{HOTAS}[HOTAS]{Hands On Throttle And Stick}
\acrodef{HUD}[HUD]{Heads Up Display}
\acrodef{PPG}[PPG]{Photoplethysmography}
\acrodef{EDA}[EDA]{Electrodermal Activity}
\acrodef{ST}[ST]{Skin Temperature}
\acrodef{ECG}[ECG]{Electrocardiogram}
\acrodef{ANS}[ANS]{Autonomic Nervous System}
\acrodef{CNS}[CNS]{Central Nervous System}
\acrodef{oop}[OOP]{Object Oriented Programming}
\acrodef{EMG}[EMG]{Electromyography}
\acrodef{sVRI}[sVRI]{Stress-Induced Vascular Response Index}

\usepackage{comment}
\usepackage{bm}
\usepackage{outlines}
\usepackage{color}
\usepackage{hyperref}
\hypersetup{pdfborder = 0 0 0}
\usepackage{microtype} 
\usepackage{tabularx}
\usepackage{algorithm,algpseudocode}
\usepackage{adjustbox}
\usepackage[rightcaption]{sidecap}
\sidecaptionvpos{figure}{t}

\usepackage{callouts}
\usepackage{tikz}
\usepackage{pgfgantt}
\usetikzlibrary{calc, arrows,arrows.meta,positioning,graphs,automata, shapes, shapes.geometric, shapes.multipart}
\usetikzlibrary{tikzmark,arrows.meta}
\tikzset{
    >=stealth',
    defnode/.style={
           rectangle,
           rounded corners,
           draw=black, thick,
           minimum height=2em,
           text centered},
    defedge/.style={
           ->,
           thick,
           shorten <=2pt,
           shorten >=2pt,}
}
\tikzstyle{component} = [draw, fill=blue!20, align=center, text centered, rounded corners]
\tikzstyle{human_stage} = [draw, fill=green!15, minimum width=5em, text centered, rounded corners]
\pgfdeclarelayer{background}
\pgfdeclarelayer{foreground}
\pgfsetlayers{background,main,foreground}
\RequirePackage{suffix}
\RequirePackage{soul}
\extrafloats{100}

\newcommand{\neweq}[2]{
\begin{equation}
 \label{e:#1}
 #2
\end{equation}
}

\WithSuffix\newcommand\neweq*[1]{
	$$
		#1
	$$
}

\usepackage{booktabs}
\usepackage{array}
\usepackage{colortbl}
\newcolumntype{L}[1]{>{\raggedright\let\newline\\\arraybackslash\hspace{0pt}}m{#1}}
\newcolumntype{C}[1]{>{\centering\let\newline\\\arraybackslash\hspace{0pt}}m{#1}}
\newcolumntype{R}[1]{>{\raggedleft\let\newline\\\arraybackslash\hspace{0pt}}m{#1}}
\newcolumntype{N}{@{}m{0pt}@{}}
\RequirePackage{float}
\restylefloat{table}
\newcommand{\newtable}[5][\normalsize]{
\setcounter{oli}{0}
\let\backupone\1
\let\backuptwo\2
\let\1\tablevelone
\let\2\tableveltwo
\begin{table}
	\begin{center}
		\caption{#4}
		#1
		\begin{tabular}{#3}
			#5
		\end{tabular}
		\label{t:#2}
	\end{center}
\end{table}
\let\1\backupone
\let\2\backuptwo
}

\makeatletter

\newcounter{oli}
\newcounter{olii}[oli]
\providecommand\theleveli{\@arabic\c@oli}%
\providecommand\thelevelii{\theleveli.\@arabic\c@olii}%
\makeatother

\providecommand\tablevelone{\rowcolor{gray!25}\setcounter{olii}{0}\refstepcounter{oli}\theleveli}
\providecommand\tableveltwo{\refstepcounter{olii}\thelevelii}

\RequirePackage{amsfonts}
\RequirePackage{amsmath}

\RequirePackage{graphicx}
\RequirePackage[verbose]{wrapfig}
\RequirePackage{float}
\RequirePackage{subfig}

\newcommand{\tikzfig}[4]{
\begin{figure*}[#2]

	#4
	\vspace{-1em}
	\caption{#3}
	\label{f:#1}
\end{figure*}
}

\newcommand{\tikzfigcol}[3]{
\begin{figure}[!t]
    \raggedright
	#3
	\vspace{-1em}
	\caption{#2}
	\label{f:#1}
    
\end{figure}
}

\def\fps@figure{htbp}
\def\fps@table{htbp}

\makeatletter

\newcounter{hypo}
\newcounter{subhypo}[hypo]
\makeatother

\definecolor{lgreen} {RGB}{180,210,100}
\definecolor{dblue}  {RGB}{20,66,129}
\definecolor{ddblue} {RGB}{11,36,69}
\definecolor{lred}   {RGB}{220,0,0}
\definecolor{nred}   {RGB}{224,0,0}
\definecolor{norange}{RGB}{230,120,20}
\definecolor{nyellow}{RGB}{255,221,0}
\definecolor{ngreen} {RGB}{98,158,31}
\definecolor{dgreen} {RGB}{78,138,21}
\definecolor{nblue}  {RGB}{28,130,185}
\definecolor{jblue}  {RGB}{20,50,100}

\definecolor{GreenYellow}       {RGB}{217, 229, 6} 	    
\definecolor{Yellow}            {RGB}{254, 223, 0} 	    
\definecolor{Goldenrod}         {RGB}{249, 214, 22} 	
\definecolor{Dandelion}         {RGB}{253, 200, 47} 	
\definecolor{Apricot}           {RGB}{255, 170, 123} 	
\definecolor{Peach}             {RGB}{255, 127, 69} 	
\definecolor{Melon}             {RGB}{255, 129, 141} 	
\definecolor{YellowOrange}      {RGB}{240, 171, 0} 	    
\definecolor{Orange}            {RGB}{255, 88, 0} 	    
\definecolor{BurntOrange}       {RGB}{199, 98, 43} 	    
\definecolor{Bittersweet}       {RGB}{189, 79, 25} 	    
\definecolor{RedOrange}         {RGB}{222, 56, 49} 	    
\definecolor{Mahogany}          {RGB}{152, 50, 34} 	    
\definecolor{Maroon}            {RGB}{152, 30, 50} 	    
\definecolor{BrickRed}          {RGB}{170, 39, 47} 	    
\definecolor{Red}               {RGB}{255, 0, 0}        
\definecolor{BrilliantRed}      {RGB}{237, 41, 57} 	    
\definecolor{OrangeRed}         {RGB}{231, 58, 0} 	    
\definecolor{RubineRed}         {RGB}{202, 0, 93}       
\definecolor{WildStrawberry}    {RGB}{203, 0, 68} 	    
\definecolor{Salmon}            {RGB}{250, 147, 171} 	
\definecolor{CarnationPink}     {RGB}{226, 110, 178} 	
\definecolor{Magenta}           {RGB}{255, 0, 144} 	    
\definecolor{VioletRed}         {RGB}{215, 31, 133} 	
\definecolor{Rhodamine}         {RGB}{224, 17, 157} 	
\definecolor{Mulberry}          {RGB}{163, 26, 126} 	
\definecolor{RedViolet}         {RGB}{161, 0, 107} 	    
\definecolor{Fuchsia}           {RGB}{155, 24, 137} 	
\definecolor{Lavender}          {RGB}{240, 146, 205} 	
\definecolor{Thistle}           {RGB}{222, 129, 211} 	
\definecolor{Orchid}            {RGB}{201, 102, 205} 	
\definecolor{DarkOrchid}        {RGB}{153, 50, 204} 	
\definecolor{Purple}            {RGB}{182, 52, 187} 	
\definecolor{Plum}              {RGB}{79, 50, 76} 	    
\definecolor{Violet}            {RGB}{75, 8, 161} 	    
\definecolor{RoyalPurple}       {RGB}{82, 35, 152} 	    
\definecolor{BlueViolet}        {RGB}{33, 7, 106} 	    
\definecolor{Periwinkle}        {RGB}{136, 132, 213} 	
\definecolor{CadetBlue}	  	    {RGB}{95, 158, 160} 	
\definecolor{CornflowerBlue}  	{RGB}{99, 177, 229} 	
\definecolor{MidnightBlue}	  	{RGB}{0, 65, 101} 	    
\definecolor{NavyBlue}          {RGB}{0, 70, 173}       
\definecolor{RoyalBlue}         {RGB}{0, 35, 102}       
\definecolor{Blue}              {RGB}{0, 24, 168}       
\definecolor{Cerulean}          {RGB}{0, 122, 201}      
\definecolor{Cyan}              {RGB}{0, 159, 218}      
\definecolor{ProcessBlue}       {RGB}{0, 136, 206}      
\definecolor{SkyBlue}           {RGB}{91, 198, 232}     

\definecolor{Turquoise}         {RGB}{0, 255, 239} 	    

\definecolor{TealBlue}          {RGB}{0, 124, 146} 	    
\definecolor{Aquamarine}        {RGB}{0, 148, 179} 	    
\definecolor{BlueGreen}         {RGB}{0, 154, 166} 	    
\definecolor{Emerald}           {RGB}{80, 200, 120} 	
\definecolor{JungleGreen}       {RGB}{0, 115, 99} 	    
\definecolor{SeaGreen}          {RGB}{0, 176, 146} 	    
\definecolor{Green}             {RGB}{0, 173, 131} 	    
\definecolor{ForestGreen}       {RGB}{0, 105, 60} 	    
\definecolor{PineGreen}         {RGB}{0, 98, 101} 	    
\definecolor{LimeGreen}         {RGB}{50, 205, 50} 	    
\definecolor{YellowGreen}       {RGB}{146, 212, 0} 	    
\definecolor{SpringGreen}       {RGB}{201, 221, 3} 	    
\definecolor{OliveGreen}        {RGB}{135, 136, 0} 	    
\definecolor{RawSienna}         {RGB}{149, 82, 20} 	    
\definecolor{Sepia}             {RGB}{98, 60, 27} 	    
\definecolor{Brown}             {RGB}{134, 67, 30}      
\definecolor{Tan}               {RGB}{210, 180, 140}	
\definecolor{Gray}              {RGB}{139, 141, 142} 	
\definecolor{DarkSlateGray}     {RGB}{57, 87, 86} 	    

\definecolor{CeruleanFrost}     {RGB}{106, 159, 200}     
\definecolor{PastelPink}        {RGB}{211, 162, 156}     

\definecolor{Black}		  	    {RGB}{30, 30, 30}       
\definecolor{White}		  	    {RGB}{255, 255, 255}    

\title{A Virtual Reality Simulation Pipeline for Online Mental Workload Modeling}

\author{Robert L. Wilson$^{1}$, Daniel Browne$^{1}$, Jon Wagstaff$^{1}$, Steve McGuire$^{1}$%

\thanks{$^{1}$Department of Electrical and Computer Engineering, University of California Santa Cruz. Santa Cruz, CA. 95060 email: robert.wilson@ucsc.edu}}

\abstract{Seamless human robot interaction (HRI) and cooperative human-robot (HR) teaming critically rely upon accurate and timely human mental workload (MW) models. Cognitive Load Theory (CLT) suggests representative physical environments produce representative mental processes; physical environment fidelity corresponds with improved modeling accuracy. Virtual Reality (VR) systems provide immersive environments capable of replicating complicated scenarios, particularly those associated with high-risk, high-stress scenarios. Passive biosignal modeling shows promise as a noninvasive method of MW modeling. However, VR systems rarely include multimodal psychophysiological feedback or capitalize on biosignal data for online MW modeling. Here, we develop a novel VR simulation pipeline, inspired by the NASA Multi-Attribute Task Battery II (MATB-II) task architecture, capable of synchronous collection of objective performance, subjective performance, and passive human biosignals in a simulated hazardous exploration environment. Our system design extracts and publishes biofeatures through the Robot Operating System (ROS), facilitating real time psychophysiology-based MW model integration into complete end-to-end systems. A VR simulation pipeline capable of evaluating MWs online could be foundational for advancing HR systems and VR experiences by enabling these systems to adaptively alter their behaviors in response to operator MW.
}

\CCScatlist{
  \CCScatTwelve{Human Robot Interaction}{Human Robot Teaming}{Virtual Reality}{Cognitive Load Theory}
  \CCScatTwelve{Mental Workload}{Psychophysiology}{Biosignals}{Multimodal Modeling}
}

\begin{document}

\maketitle

\section{Introduction}
\label{sec:intro}

Frameworks for \ac{MW} modeling, particularly for implementation in \ac{HRI} and teaming, typically operate on limited biosignals and are incapable of operating online, leaving room for vast improvements in \ac{MW} model development and evaluation. \ac{HR} teams, in both the physical and virtual domains, are transitioning to peer-like paradigms where each team member can dynamically respond to the current state of the other \cite{holden2013evidence,johnson2014coactive,ma2018human}. In particular, improvements in \ac{MW} for high-risk environments (e.g., driving, surgery, exploration) stand to improve safety and mission outcomes \cite{heard2017human}. However, data collection in these environments, a known strategy for improving model outcomes \cite{charles2019measuring}, can be risky and quite costly \cite{chao2017effects}. \ac{VR} is a key tool for improving \ac{MW} models through data collection closer to the operational environment, without the costs and risks associated with real operations.

Model mismatches between \ac{HR} teammates, often due to poor \ac{MW} models, can risk detrimental outcomes. Human operators function optimally within individual workload boundaries \cite{haapalainen2010psycho,vanneste2020towards,zepf2020driver}. Cognitive overload can lead to stress and operational mistakes, while cognitive underload can result in a lack of motivation or attention drift and decrease team performance. Operator interruption at inopportune times can increase cognitive demands and negatively impact mission efficacy \cite{haapalainen2010psycho}. 

\ac{CLT}, the preeminent model of human cognitive architecture, provides a theoretical framework for characterizing and formalizing \ac{MW} \cite{paas2003cognitive,paas1994instructional,sweller1998cognitive}. \ac{CLT} suggests \ac{MW} is a composite of intrinsic, extrinsic, and germane loads. Intrinsic load relates to the effort required to learn a particular concept. The section of \ac{MW} related to the information presentation format defines the extrinsic load. Germane load encompasses the processing involved in understanding information relevant to the task at hand. Different component combinations indicate various levels of \ac{MW}. The NASA \ac{MATB-II} provides a well-studied and useful structure for manipulating \ac{MW} \cite{daviaux2019feedback,santiago2011multi}. \ac{MATB-II} consists of four main sections: \ac{RM}, tracking, \ac{SM}, and communications. The difficulty of each \ac{MATB-II} section can be independently altered, allowing for fine-grained experimental control. However, the use of \ac{MATB-II} has been mostly relegated to traditional on-screen testing \cite{albuquerque2020wauc,heard2019multi}.

\ac{MW} modeling predictions could improve via a robust multimodal model based on passive biosignals integrated into a \ac{VR} system. Regardless of the chosen \ac{MW} model (e.g., supervised learning, reinforcement learning), psychophysiological analysis aims to map a hidden cognitive state (e.g., \ac{MW}) to observable external manifestations by teasing out concomitants (physical traits that map to many cognitive functions) \cite{backs2000application}. Multimodal biosensor models are on the rise, but are rarely robust in design (i.e., include more than two biosignals), allowing for limited model performance \cite{charles2019measuring,vanneste2020towards}. \ac{CLT} indicates data collection and downstream model performance are likely to improve with the use of task relevant data \cite{choi2014effects}. However, most \ac{MW} work does not employ a relevant task for data collection and training, introducing a major disconnect between theory and practice. A \ac{VR} environment not only enables studies with higher task relevancy, which is particularly useful for \ac{MW} models aimed at high-risk scenarios, but also facilitates online model use in addition to more traditional offline evaluation.

Here, we describe an online \ac{VR} environment, specifically designed to modulate \ac{MW}, and the real-time data the pipeline produces. A robust sweep of biosensors allows for multimodal feature extraction. The \ac{VR} framework, complete with realistic motion cues, enables deployment and possible evaluation of both closed (online) and open (offline) \ac{MW} models, in task relevant settings. The \ac{VR} pipeline facilitates passive biosignal generation potentially closer to those that could be elicited during the actual task, thus likely improving \ac{MW} model performance, particularly in high-risk scenarios.

\tikzfig{block}{!t}{Our lunar rover \ac{VR} simulation produces online time synced human performance and in-simulation objective measures for investigation of online mental workload estimation. The exploratory simulation, founded on \ac{MATB-II}, allows for manipulation of each component separately. All four control components play a role in a \ac{MW} model (i.e., Intrinsic, Extrinsic, Germane Load). A biosensor suite records a robust sweep of raw physiological data and extracts relevant biofeatures data during play (see \autoref{sec:biofeatures}). In-game objective data and extracted biofeatures are time synced through ROS. Between-run TLX survey results paired to time synced biofeatures and objective measures enable \ac{MW} modeling online and post simulation.\vspace{-1em}}{

\tikzstyle{component} = [draw, text centered, rounded corners,font=\normalsize, inner sep=1, outer sep=0]
\def\blockdist{1}
\def\hblockdist{.5}
\def\vblockdist{1}
\def\edgedist{1.5}

\begin{tikzpicture}[scale=1, transform shape, post/.style={->,shorten >=1pt,>=stealth',semithick, rounded corners=4pt}]

    \path (0,0) node (sim_title) [component,fill=blue!50,align=center, anchor=west,inner sep=3,text width=59ex] {\textbf{Lunar Rover Simulation Environment}};

	\path (sim_title.south west) +(0,-.425*\vblockdist) node (pupil) [component, fill=blue!30, anchor=north west,text width=18ex,text depth=10ex] {
	\begin{tabular}{c}
	Cardiovascular \\
	\end{tabular}
	\adjincludegraphics[Clip={{.25\width} {.05\height} {.25\width} {.05\height}}, width=.95\textwidth]{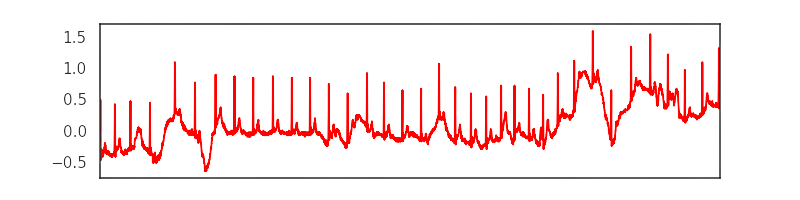}
	};
	
	\path (pupil.south west) +(0,-.15*\vblockdist) node (resp) [component, fill=blue!30, anchor=north west,text width=18ex,text depth=10ex] {
    \begin{tabular}{c}
	Dermal Activity\\
	\end{tabular}
	\adjincludegraphics[Clip={{.25\width} {.05\height} {.25\width} {.05\height}}, width=.95\textwidth]{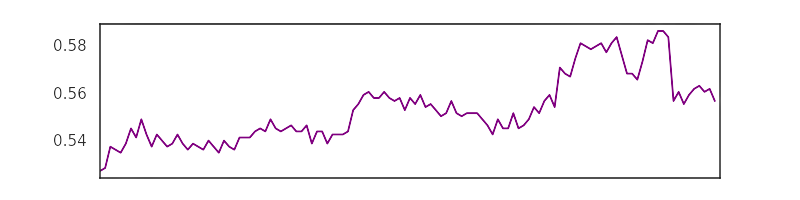}
	};
	
	\path (resp.south west) +(0,-.15*\vblockdist) node (eda) [component, fill=blue!30, anchor=north west,text width=18ex,text depth=10ex] {
		\begin{tabular}{c}
	Respiration\\
	\end{tabular}
	\adjincludegraphics[Clip={{.25\width} {.05\height} {.25\width} {.05\height}}, width=.95\textwidth]{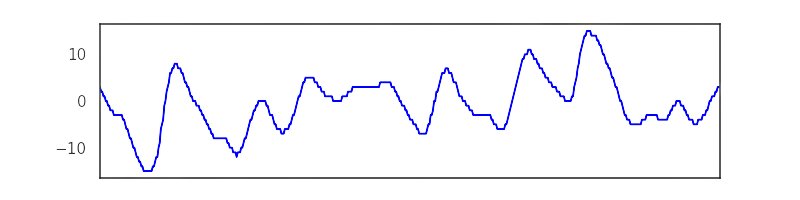}
	};
	
	\path (eda.south west) +(0,-.15*\vblockdist) node (heart) [component, fill=blue!30, anchor=north west, text width=18ex,text depth=10ex] {
	\begin{tabular}{c}
	Pupillometry \\
	\end{tabular}
	\adjincludegraphics[Clip={{.27\width} {.05\height} {.25\width} {.05\height}}, width=.95\textwidth]{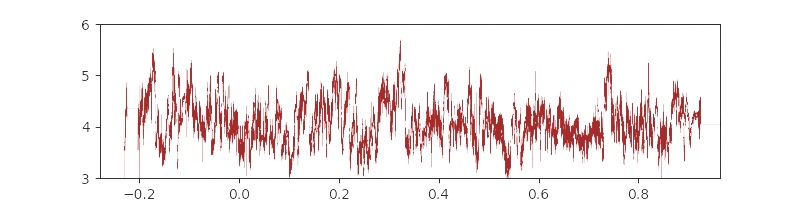}
	};
	
	\path (pupil.north east -| heart.east)+(.5*\hblockdist,0) node (task) [component, fill=Gray!30, anchor= north west, text width=40ex,text depth=50.5ex] {
	\begin{tabular}{c}
	VR Exploration Task
	\end{tabular}
	\includegraphics[width=.95\textwidth]{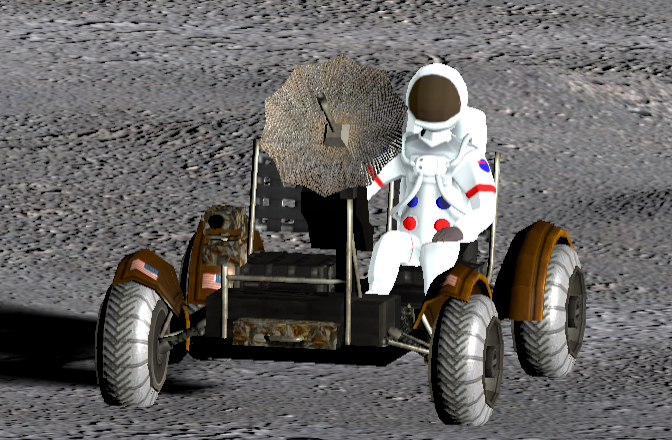}
    \adjincludegraphics[width=1.12\textwidth,height=.58\textwidth,Clip={{.15\width} {0\height} {.01\width} {0\height}} ]{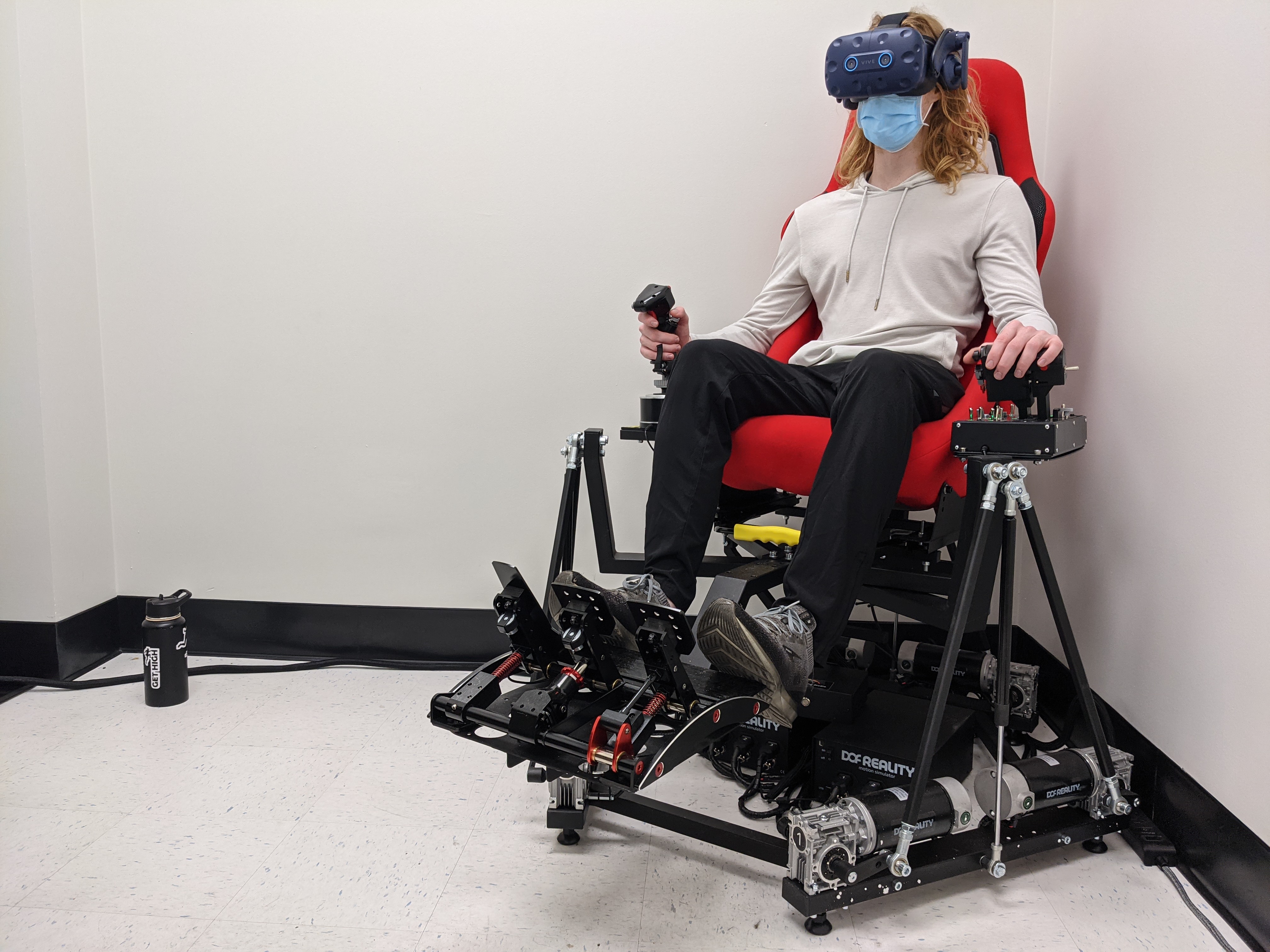}
	};
    
	\draw  ($ (resp.south west)!0.5!(eda.north west) $)+(-1.25*\hblockdist,0) node [draw, circle, x radius=.5em, y radius=.5em] (bind) {};

	\begin{pgfonlayer}{background}
	\def\borderwidth{0.3}
    \path (sim_title.south west -| bind.north)+(0,\borderwidth) node (sim_a) {};
    \path (task.south east)+(\borderwidth,-\borderwidth) node (sim_b) {};
    \path[fill=blue!15,rounded corners, draw=black!50, dashed] 
        (sim_a) rectangle (sim_b) node (sim) {};

    \end{pgfonlayer}

	\path (sim_title.north west) + (-1.5*\hblockdist ,-.125*\vblockdist) node (controls) [component,fill=Orange!30, anchor = north east] {
	\begin{tabular}{c} 
	\underline{MATB-II Style Controls}\\
	Tracking\\
	System Monitoring\\
	Resource Management\\
	Communications
	\end{tabular}};
	
	\path (controls.south) + (0, -1.25*\vblockdist) node (processes) [component,fill=Orange!30] {
	\begin{tabular}{c} 
	\underline{Mental Workload}\\
	Intrinsic Load\\
	Extrinsic Load\\
	Germane Load
	\end{tabular}};

	\path (processes.south |- bind.south) + (0, -2*\vblockdist) node (markers) [component,fill=Orange!30,align=center, text width=23ex, text depth=10ex] {
    \begin{tabular}{m{8ex} c} 
	    \multicolumn{2}{c}{\underline{Biosensor Suite}}\\
	    \centering \textit{\footnotesize Empatica E4}\vspace{5ex}&
	    \includegraphics[width=.3\textwidth]{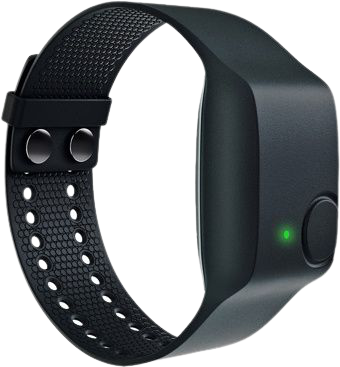}\vspace{-3ex}\\
	    \centering \textit{\footnotesize Zephyr BH}\vspace{5ex}&
	    \includegraphics[width=.5\textwidth]{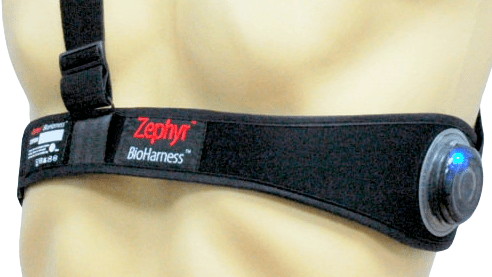}\vspace{-3ex}\\

	    \centering \textit{\footnotesize Vive Pro Eye}\vspace{5ex}&
	    \includegraphics[width=.45\textwidth]{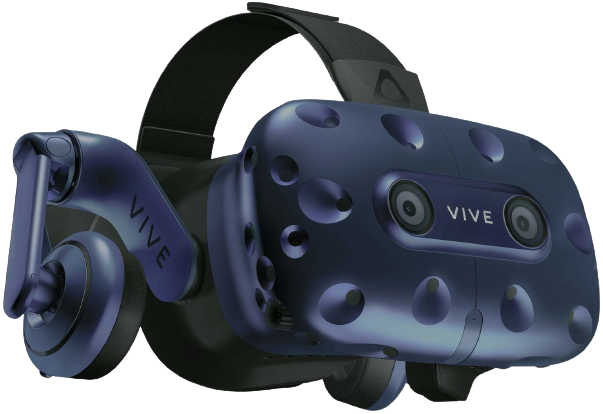}\\
    \end{tabular}

	};


	\path (task.north east)+(3*\hblockdist,.25*\vblockdist)  node (objmeasures) [component, fill=yellow!30, anchor=north west] {
	\begin{tabular}{c}Objective\\Performance\\Measures
	\end{tabular}};	
	
	\path (objmeasures.south)+(0,-.15*\vblockdist) node (subjmeasures) [component, fill=yellow!30, anchor=north] {
	\begin{tabular}{c}Subjective\\TLX\\Survey
	\end{tabular}};	
	
	\path (subjmeasures.south)+(0,-.15*\vblockdist) node (human_gen) [component, fill=yellow!30, anchor=north] {
	\begin{tabular}{c}Derived\\Biosignal\\Features
	\end{tabular}};

	\begin{pgfonlayer}{background}
	\def\borderwidth{0.3}
    \path (objmeasures.north -| objmeasures.west)+(-\borderwidth,\borderwidth) node (output_a) {};
    \path (human_gen.south -| objmeasures.east)+(\borderwidth,-\borderwidth) node (output_b) {};
    \path[fill=yellow!15,rounded corners, draw=black!50, dashed]
        (output_a) rectangle (output_b);
     \end{pgfonlayer}

	\draw  ($ (output_a)!0.50!(output_b) $) + (0, -2.15*\vblockdist) node [draw, circle] (bind2) {}; 
	
	 \path ($ (output_a)!0.50!(output_b) $) + (0, 2.15*\vblockdist) node (out_title) [component,fill=yellow!50,align=center,inner sep=2] {\textbf{Online Data Collection}};

	\path  ($ (output_a)!0.50!(output_b) $) + (0,-4.25*\vblockdist) node (ml) [component, fill=green!20, anchor=south] {
	
	\begin{tabular}{c}
	\underline{Mental Workload Models}\\Multivariate Linear\\Supervised Learning\\Reinforcement Learning
	\end{tabular}
	
	};
	
	\draw[post] (controls.south)  -- (processes.north);
	\path (bind.west)+(.3*\hblockdist,.07\vblockdist) node (bind_topleft){};
	\draw[post] (processes.south)  |- (bind_topleft);
	\path (bind.west)+(.3*\hblockdist,-.07\vblockdist) node (bind_botleft){};
	\draw[post] (markers.north)  |- (bind_botleft);
	\draw[post] ($ (controls.north east)!0.275!(controls.south east) $)  -| (task.north);

	\draw[post] (bind.east) |- (pupil.west);
	\draw[post] (bind.east) |- (heart.west);
	\draw[post] (bind.east) |- (eda.west);
	\draw[post] (bind.east) |- (resp.west);

	\draw[post] (pupil.east)  -- (task.west |- pupil.east);
    \draw[post] (resp.east)  -- (task.west |- resp.east);
	\draw[post] (eda.east)  -- (task.west |- eda.east);
	\draw[post] (heart.east)  -- (task.west |- heart.east);
	
	\path ($ (task.south)!0.5!(task.south west) $)+(0,.15*\vblockdist) node (task_bl2){};
	
	\path (task.east)+(0,.287*\vblockdist) node (task_rt){};
    
	\path ($ (task.south)!0.5!(task.south east) $)+(0,.15*\vblockdist) node (task_br){};
	
    \path (ml.north)+(-1*\hblockdist,-.15*\vblockdist) node (ml_lt){};
    
    \draw[post] (bind2.south)  -- (ml.north);

    \path (objmeasures.west) + (-0.4*\hblockdist,0) node (data_l){};
    \draw[post] (task.east |- objmeasures.west) -- (data_l);
	\draw[post,dashed,color=black] ($ (ml.south)!0.5!(ml.south east) $) -- +(0,-1*\vblockdist) -- +(2*\hblockdist,-1*\vblockdist);

    \path ($ (ml.south)!0.5!(ml.south east) $) -- +(1*\hblockdist,-.5*\vblockdist) node (open_text) {
	\begin{tabular}{c}
	\textit{Open} \\
	\textit{Loop}
	\end{tabular}
	};
	
	\draw[post,dashed,color=black] ($ (ml.south)!0.5!(ml.south west) $) -- +(0,-1*\vblockdist) -- +(-3.05*\hblockdist,-1*\vblockdist);

    \path ($ (ml.south)!0.5!(ml.south west) $) -- +(-1.25*\hblockdist,-.5*\vblockdist) node (close_text) {
	\begin{tabular}{c}
	\textit{Closed} \\
	\textit{Loop}
	\end{tabular}
	};

\end{tikzpicture}
}

\section{Relevant Background}
\label{sec:related}

We constructed our \ac{VR} model by combining multiple aspects of research in both \ac{HRI} and psychology into a multifaceted approach. Performance measures, both objective and subjective, enable external measurement of human achievement. Biosignals allow inference of the \ac{ANS} and \ac{CNS}, potentially providing deeper insights. \ac{VR} is a novel test bed for \ac{MW} studies promoting task relevant simulations.

\subsection{Performance Measures}

Performance measures display an intricate relationship to \ac{MW} \cite{galy2012relationship}. Optimum \ac{MW}s consistently yield maximized objective performance \cite{heard2019multi,vanneste2020towards}. In \ac{VR}, objective performance measures have demonstrated improved task performance and been linked to \ac{MW} \cite{bailenson2008effect,tichon2006training}. A multitude of methods have been subjective surveys have been explored in relation to \ac{MW} \cite{rubio2004evaluation}. The \ac{TLX} survey has demonstrated a relationship to \ac{MW} in a variety of modalities \cite{hart2006nasa}, including proving the usefulness of \ac{VR} in modulating \ac{MW} \cite{chao2017effects,pouliquen2016remote}. 

\tikzfig{hud}{!b}{The \ac{VR} simulation is founded on \ac{MATB-II} enabling easy manipulation of each characteristic. (A) The user must navigate the lunar terrain to an objective point. (B) Updating the radar position with respect to a fixed location requires the user concentrate on tracking. (C) Switching radio communications channels, prompted by separate voices, aims to satisfy the communications requirement. (D) The different rates of O$_2$ usage and CO$_2$ buildup force the user to consider resource management. (E) Motor temperature fluctuations and limitations require system management. Together, all pieces of the simulation act to modulate \ac{MW} based on in-simulation difficulty.\vspace{-1em}}{

\tikzstyle{component} = [ text centered, rounded corners,font=\Large, inner sep=2, outer sep=0, fill=Goldenrod, anchor=north west, text opacity=1, text=black]
\def\blockdist{1}
\def\hblockdist{.5}
\def\vblockdist{1}

\begin{tikzpicture}[scale=1, transform shape, post/.style={->,shorten >=1pt,>=stealth',semithick, rounded corners=4pt}]

    
    \path (0,0) node (hud) [align=left] {\includegraphics[width=\textwidth, height = 7.5cm ]{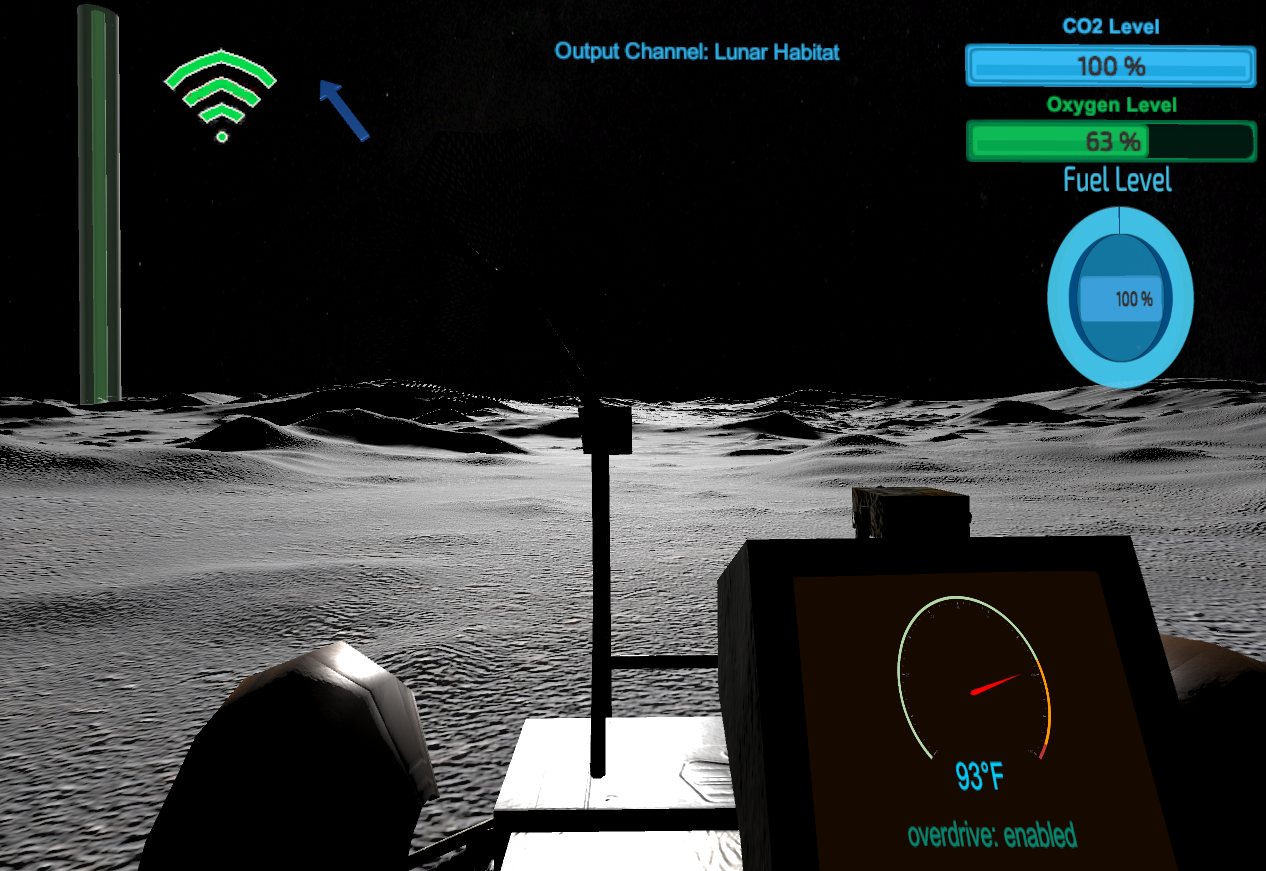}};
    
    \path
        let
            \p1 = ($(hud.west)!0.125!(hud.east)$),
            \p2 = ($(hud.north)!0.05!(hud.south)$),
            \p3 = ($(hud.west)!0.31!(hud.east)$),
            \p4 = ($(hud.north)!0.205!(hud.south)$),
            \p5 = ($(hud.west)!0.055!(hud.east)$),
            \p6 = ($(hud.north)!0.02!(hud.south)$),
            \p7 = ($(hud.west)!0.112!(hud.east)$),
            \p8 = ($(hud.north)!0.48!(hud.south)$),
            \p9 = ($(hud.west)!0.43!(hud.east)$),
            \p{10} = ($(hud.north)!0.05!(hud.south)$),
            \p{11} = ($(hud.west)!0.67!(hud.east)$),
            \p{12} = ($(hud.north)!0.1!(hud.south)$),
            \p{13} = ($(hud.west)!0.755!(hud.east)$),
            \p{14} = ($(hud.north)!0.02!(hud.south)$),
            \p{15} = ($(hud.west)!0.991!(hud.east)$),
            \p{16} = ($(hud.north)!0.46!(hud.south)$),
            \p{17} = ($(hud.west)!0.695!(hud.east)$),
            \p{18} = ($(hud.north)!0.65!(hud.south)$), 
            \p{19} = ($(hud.west)!0.86!(hud.east)$), 
            \p{20} = ($(hud.north)!0.98!(hud.south)$), 
        in
        coordinate (p1) at (\p1)
        
        coordinate (p2) at (\p2)
        coordinate (p3) at (\p3)
        coordinate (p4) at (\p4)
       
        coordinate (p5) at (\p5)
        coordinate (p6) at (\p6)
        coordinate (p7) at (\p7)
        coordinate (p8) at (\p8)
        
        coordinate (p9) at (\p9)
        coordinate (p10) at (\p{10})
        coordinate (p11) at (\p{11})
        coordinate (p12) at (\p{12})
        
        coordinate (p13) at (\p{13})
        coordinate (p14) at (\p{14})
        coordinate (p15) at (\p{15})
        coordinate (p16) at (\p{16})
        
        coordinate (p17) at (\p{17})
        coordinate (p18) at (\p{18})
        coordinate (p19) at (\p{19})
        coordinate (p20) at (\p{20});

    \path[rounded corners, draw=white, dashed, thick] (p1 |- p2) rectangle (p3 |- p4)  node (radar) {};
    
    \path[rounded corners, draw=white, dashed, thick] (p5 |- p6) rectangle (p7 |- p8)  node (sci) {};
    
    \path[rounded corners, draw=white, dashed, thick] (p9 |- p10) rectangle (p11 |- p12)  node (radio) {};
    
    \path[rounded corners, draw=white, dashed, thick] (p13 |- p14) rectangle (p15 |- p16)  node (levels) {};
    
    \path[rounded corners, draw=white, dashed, thick] (p17 |- p18) rectangle (p19 |- p20)  node (motor) {};

    \path ($(p1 |- p4) !0.5! (p3 |- p4)$)  node (rad_mid) [component,anchor=center, fill opacity=0] {};
    
    \path ($(p5 |- p8) !0.5! (p7 |- p8)$)  node (obj_mid) [component,anchor=center, fill opacity=0] {};
    
    \path ($(p9 |- p12) !0.5! (p11 |- p12)$)  node (com_mid) [component,anchor=center, fill opacity=0] {};
    
    \path ($(p13 |- p14) !0.5! (p13 |- p16)$)  node (rm_lmid) [component,anchor=center, fill opacity=0] {};
    
    \path ($(p17 |- p18) !0.5! (p17 |- p20)$)  node (sm_lmid) [component,anchor=center, fill opacity=0] {};
    
    \path (rad_mid)+(0,-.5*\vblockdist)  node (track) [component, anchor = north, fill opacity=.7] {B.) Tracking};

    \path (obj_mid)+(0,-.5*\vblockdist)  node (obj) [component, anchor = north, fill opacity=.7] {A.) Objective};

    \path (com_mid)+(0,-.5*\vblockdist)  node (comms) [component, anchor = north, fill opacity=.7] {C.) Communications};

    \path (rm_lmid)+(-\hblockdist,-\hblockdist)  node (resources) [component, anchor =north east, fill opacity=.7] {
    \begin{tabular}{c}
         D.) Resource\\
         Management
    \end{tabular}};

    \path (sm_lmid)+(-\hblockdist,0)  node (sys) [component, anchor = east, fill opacity=.7] {
    \begin{tabular}{c}
         E.) System\\
         Management
    \end{tabular}};

    \draw[post,color=white] (track.north)  -- (rad_mid);
    
    \draw[post,color=white] (obj.north)  -- (obj_mid);
    
    \draw[post,color=white] (comms.north)  -- (com_mid);
    
    \draw[post,color=white] (resources.north east)  -- (rm_lmid);
    
    \draw[post,color=white] (sys.east)  -- (sm_lmid);

\end{tikzpicture}
    \vspace{-1em}
}

\subsection{Biosignals}
\label{sec:previous_biosignals}

Physiological signals give quantitative insight into an individual \ac{MW} \cite{charles2019measuring}. For our work, we selected six raw biosignals (and multiple derived features) that have demonstrated  \ac{MW} correlations: \ac{ECG}, \ac{EDA}, \ac{PPG}, Respiration, \ac{ST}, and Pupillometry. 

\ac{ECG}, derived from cardiac electrical activity, and \ac{PPG}, blood flow interpolation via light refraction, provide indirect insight into \ac{MW} via cardiovascular activity measurement \cite{charles2019measuring}. Temporally, \ac{ECG} measures (\ac{HR} and \ac{HRV}) have demonstrated relationships to \ac{MW} \cite{charles2019measuring, rieger2014heart, vanneste2020towards}. In the frequency domain, power bands increase or decrease as a function of task load \cite{charles2019measuring}. \ac{PPG}, either individually or in conjunction with \ac{ECG}, may provide more useful \ac{MW} estimates \cite{wang2019data}. For example, calculating \ac{sVRI}, a \ac{PPG} signal derivative, requires a lower time period of analysis to estimate \ac{MW} than \ac{HRV}, making it a good candidate for real-time estimation \cite{zhang2018evaluating}. Notably, physical exertion confounds cardiovascular measures, making them poorly suited for tasks with significant physical loads \cite{grassmann2016respiratory}. 
 
Dermal measures relate to \ac{MW} as a response to general \ac{ANS} stimulation. \ac{EDA}, also known as \ac{GSR}, quantifies the electrical resistance of skin. \ac{EDA} waveforms consist of tonic, slow acting, and phasic, event specific, responses \cite{braithwaite2013guide}. Tonic components have indicated moderate utility for \ac{MW} estimation (e.g., tonic components loosely correlate with \ac{MW}), while shorter, more frequent, phasic peaks suggest a stronger \ac{MW} relationship \cite{braithwaite2013guide}. \ac{ST} can be utilized in \ac{MW} estimations, but with temporal limitations of slow rise times and delayed event responses. Generally, \ac{ST} decreases in response to \ac{MW} \cite{larmuseau2020multimodal}. However, \ac{ST} responses to \ac{MW} may be location dependent \cite{abdelrahman2017cognitive}.

The Respiration Waveform gives insight into \ac{MW} through breathing rate and variations \cite{grassmann2016respiratory}. In a relaxed state, a human breathes slowly and consistently. However, as \ac{MW} rises, overall breathing rate increases along with an increased prevalence of irregular rhythms, quick variations, and cessations \cite{koelstra2011deap}. 

Pupillometry measures fluctuations in pupil size and reactivity, which have a direct relationship to \ac{MW} through the \ac{CNS} \cite{marquart2015review}. For example, variations in the rate and magnitude of microsaccades, involuntary eye movements that occur during fixation, can separate different \ac{MW} levels \cite{krejtz2018eye}. One study specifically records pupillometry in \ac{VR} and found a positive correlation between pupil diameter and subjective task load scores \cite{schwalm2008pupillometry}.

\subsection{VR in MW modeling}

\ac{VR} provides an immersive environment capable of simulating complex situations in a controlled manner \cite{chao2017effects}. While \ac{VR} has been employed in a variety of use cases \cite{kaufeld2019level,luong2021survey}, the utility of \ac{VR} starts to shine in high-risk scenarios. Driving simulations commonly vary route options or obstacle difficulty in order to induce different \ac{MW}s \cite{abdurrahman2021effects, imhoff2016driving, zepf2020driver}. Users of surgical simulators have subjectively reported the utility of \ac{VR} in operations preparation \cite{barre2019virtual} and objectively demonstrated decreased \ac{MW}s after system use \cite{zheng2012workload}. The use of \ac{VR} for flight simulation assessment has been able to differentiate between low and high \ac{MW}s \cite{andrievskaia2020neural, kakkos2019mental}.

\section{System Design}
\label{sec:methods}

Motivated by the cognitive state insights provided by performance and biosignal measures as well as the role \ac{VR} can play (see \autoref{sec:related}), we designed a \ac{VR} exploration environment with the aim of modulating \ac{MW} through simulation difficulty. While playing the simulation, the user dons three biosensors (i.e., Zephyr Bioharness, Empatica E4, HTC Vive Pro Eye). The system collects objective performance and physiological measures online, with \ac{TLX} survey responses collected between runs (see \autoref{f:block}). The Zephyr Bioharness communicates directly with \ac{ROS} while the Empatica E4 and Unity connect to \ac{ROS} through Windows via the vendor provided streaming server and \textit{rosbridge} respectively. \ac{ROS} time synchronizes all collected data, which can be used in a variety of \ac{MW} models, enabling real-time (closed loop) use while still providing traditional offline (open loop) analysis.

The simulation begins with the user randomly placed on a 1 km$^2$ lunar texture, onboard the Apollo rover. A prompt informs the user of the mission scenario and objective. While on an exploration mission to an area of scientific interest (i.e., objective point), the rover's autonomous systems have failed. The rover's responsibilities now fall on the human as the user navigates to the objective point.

\subsection{Simulation Design}
\label{sec:game}

The simulation tasks the user with navigating to a randomly generated objective point (represented as a green column (\autoref{f:hud}A)) and drop a marker while aboard the Apollo lunar rover, in the shortest amount of time. A 6-axis motion platform provides motion cues (\autoref{f:block}), generated by the lunar texture the user drives across, complete with corresponding lunar physics (e.g., gravity of 0.166G). A side mounted \ac{HOTAS}, in the same layout as the in simulation rover, provides in game movement and interaction (\autoref{f:block}). While navigating, the user must track or respond to various rover alerts, all chosen in relationship to the \ac{MATB-II} due to its ability to regulate the various components of \ac{MW}. Upon objective completion, a \ac{TLX} prompt appears.

\subsubsection{Tracking}

A radar indicator (\autoref{f:hud}B) details the signal strength and optimal direction of the on-board radar dish. The user must maintain optimal connection to the stationary relay station as they navigate through a button on the \ac{HOTAS} stick. The user's field of view includes the actual on-board radar for added visual feedback. As signal strength worsens, the radar indicator drops bars and changes colors (i.e., green to yellow to red), while the arrow indicates the necessary directional change. If the user does not correct a drop in signal strength, the indicator will start flashing and increase flashing frequency over time. 

\subsubsection{Communication}

While navigating, contact must be maintained with either the base station or the lunar habitat. Separate audio (i.e., male for base station and female for lunar habitat) and visual (\autoref{f:hud}C) prompts tell the user which frequency to broadcast to. Upon receiving the prompt, the user must switch channels via a button located on the \ac{HOTAS} stick. Delays in response cause the prompt to reoccur at an increased frequency. The system may also separately ask the user to report in, prompting a separate trigger response.

\subsubsection{Resource Management}

Distance traveled as well as motor speed and temperature dictate the rover's limited battery, part of the \ac{RM} triad (\autoref{f:hud}D). Overdrive capabilities, shown in the on-board display (\autoref{f:hud}E), can increase speed for a short burst but at the cost of significant battery life. Depletion of rover battery results in a failed run. CO$_2$ management, which requires the user to expel its contents after reaching a certain threshold, and O$_2$ consumption, the depletion of which results in a failed run, act as the other two constituents. Breathing rate, supplied by the Zephyr Bioharness, drives both CO$_2$ buildup and O$_2$ consumption, but at different scales (i.e., CO$_2$ buildup happens multiple times a run while O$_2$ depletion is a one time event).

\tikzfigcol{trial}{A full trial consists of four separate runs, at alternating intensities, separated by three minutes of free play. Time synchronized biometrics (e.g., pupil diameter), objective performance (e.g., motor temperature), and subjective (\ac{TLX} survey administration) performance measures enable online information use. At run initialization, the user must navigate to the objective point, while all metrics are recorded in real time. Periods of free play separate each run allowing the user biosignals to return to baseline. \vspace{-1em}}{

\tikzstyle{component} = [text centered, rounded corners,font=\Large, inner sep=0, outer sep=0, fill=Goldenrod, anchor=north west, text opacity=1, text=black]
\def\blockdist{1}
\def\hblockdist{1}
\def\vblockdist{.10}
\def\recfill{.3}

\def\imsz{1\columnwidth}
\def\botclip{0.1}
\def\lftclip{0.229}
\def\rgtclip{0.225}
\def\topclip{0.15}

\def\toplft{0}
\def\toptop{-0.01}

\def\botlft{0}
\def\botbot{0.05}

\def\trialblock{2} 

\begin{tikzpicture}[node distance=0em, scale=1, transform shape, post/.style={->,shorten >=1pt,>=stealth',semithick}]


    \path (0,0) node (pupil) [align=left]  {\adjincludegraphics[scale={.68}{0.45},Clip={{\lftclip\width} {\botclip\height} {\rgtclip\width} {\topclip\height}}]{figures/pupilDiam.png}};
    
    \path (pupil.south west) + (0, \vblockdist) node (obj) [align=left, anchor = north west] {\adjincludegraphics[scale={.68}{0.45},Clip={{\lftclip\width} {\botclip\height} {\rgtclip\width} {\topclip\height}}]{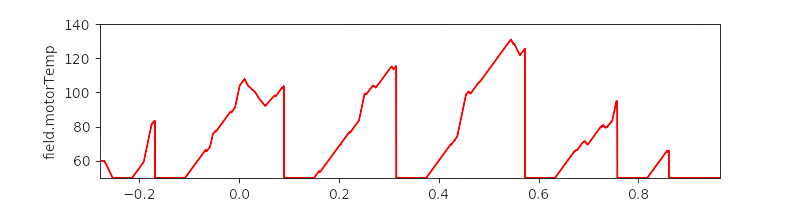}};
    
    \path (obj.south west) + (0, \vblockdist) node (subj) [align=left, anchor = north west] {\adjincludegraphics[scale={.68}{0.45},Clip={{\lftclip\width} {.115\height} {\rgtclip\width} {\topclip\height}}]{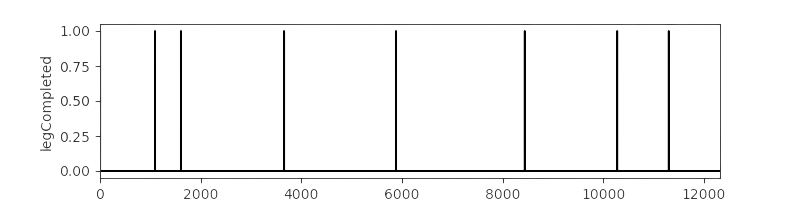}};
    
    \begin{pgfonlayer}{background}
    \path (0,0) node (org) {};
    \path (pupil.north west) + (0.05,-.13) node (ses1_tl) {};
    \path (subj.south west) + (1.87*\hblockdist,.13) node (ses1_br) {};
    
    \path (ses1_tl) + (2.3*\hblockdist,0) node (ses2_tl) {};
    \path (ses1_br) + (1.95*\hblockdist,0) node (ses2_br) {};
    
    \path (ses2_tl) + (1.95*\hblockdist,0) node (ses3_tl) {};
    \path (ses2_br) + (2.25*\hblockdist,0) node (ses3_br) {};
    
    \path (ses3_tl) + (2.15*\hblockdist,0) node (ses4_tl) {};
    \path (ses3_br) + (1.6*\hblockdist,0) node (ses4_br) {};

    \path[fill opacity=\recfill, fill=GreenYellow] (ses1_tl |- ses1_br) rectangle (ses1_tl -| ses1_br)  node (t1) {};
    \path[fill opacity=\recfill, fill=GreenYellow] (ses2_tl |- ses2_br) rectangle (ses2_tl -| ses2_br)  node (t2) {};
    \path[fill opacity=\recfill, fill=GreenYellow] (ses3_tl |- ses3_br) rectangle (ses3_tl -| ses3_br)  node (t3) {};
    \path[fill opacity=\recfill, fill=GreenYellow] (ses4_tl |- ses4_br) rectangle (ses4_tl -| ses4_br)  node (t4) {};
    
    \path ($(ses1_tl |- ses1_tl) !0.5! (ses1_br |- ses1_tl)$) + (0,.13) node (rad_mid) [font = \scriptsize ,anchor=center] {Trial I};
    \path ($(ses2_tl |- ses2_tl) !0.5! (ses2_br |- ses2_tl)$) + (0,.13) node (rad_mid) [font = \scriptsize ,anchor=center] {Trial II};
    \path ($(ses3_tl |- ses3_tl) !0.5! (ses3_br |- ses3_tl)$) + (0,.13) node (rad_mid) [font = \scriptsize ,anchor=center] {Trial III};
    \path ($(ses4_tl |- ses4_tl) !0.5! (ses4_br |- ses4_tl)$) + (0,.13) node (rad_mid) [font = \scriptsize ,anchor=center] {Trial IV};
    \end{pgfonlayer}

    \draw[color=white] ($(subj.south west)+(.2,0.14)$)  -- ($(subj.south east)+(-.1,0.14)$);
    \path ($(subj.south west)+(.15,0.15)$) node (start) [anchor=north] {0};
    \path ($(subj.south east)+(-.25,0.15)$) node (end) [anchor=north] {60};
    \path ($(start) !0.5! (end)$) node (time)[] {Time (min)};

    \path (pupil.north west) +(\toplft,\toptop) node (toppup) [anchor=north] {6};
    \path (pupil.south west) +(\botlft,\botbot) node (botpup) [anchor=south] {3};
    \path ($(toppup) !0.5! (botpup)$) + (-.3,0) node (pupunit)[rotate=90] {\begin{tabular}{c}
        \scriptsize Diameter \\
         (mm)
    \end{tabular}
    };
    
    \path (obj.north west) +(-.2,\toptop) node (topobj) [anchor=north] {140};
    \path (obj.south west) +(-.1,\botbot) node (botobj) [anchor=south] {50};
    \path ($(topobj) !0.5! (botobj)$) + (-.2,0) node (objunit)[rotate=90] {\begin{tabular}{c}
          Temp \\
         ($^\circ$C)
    \end{tabular}
    };
    
    \path (subj.north west) +(\toplft,\toptop) node (topsubj) [anchor=north] {1};
    \path (subj.south west) +(\botlft,\botbot) node (botsubj) [anchor=south] {0};
    \path ($(topsubj) !0.5! (botsubj)$) + (-.25,0) node (subjunit)[rotate=90] {\begin{tabular}{c}
          \scriptsize Amplitude \\
    \end{tabular}
    };
    
    \end{tikzpicture}
    \vspace{-1em}
    }
\subsubsection{System Monitoring}

Motor temperature and rover speed fulfill the role of \ac{SM} (\autoref{f:hud}E). The on-board display reflects rover temperature variations, which are controlled by the \ac{HOTAS} throttle on the motion platform. Throttle speed and terrain dictate the actual rover speed, with the user feeling all motion in real time through the motion platform. Vehicle torque drives motor temperature. If motor temperature reaches a critical temperature, the entire rover will stall while the system cools down, drastically affecting completion time. An overdrive button increases the torque over the recommended limit for a short time at the cost of an increased motor temperature.

\subsection{Trial Design}
\label{sec:trial}

We structured each run in a repeated measures fashion, designed to last for one-hour (\autoref{f:trial}). Each trial begins by equipping the user with biosensors and initializing the \ac{ROS} data collection. Upon situating themselves in the motion platform and being briefed on the mission objective, the user explores the lunar terrain in free play mode, with the \ac{HUD} and on-board display turned off, for five minutes. The five-minute period acts as an establishment of physiological baseline \cite{heard2019multi}. Afterward, the system sets the run's in-game difficulty, an objective point appears, and the information systems (i.e., the \ac{HUD} and on-board display) turn on. The user then completes the task as previously described (see \autoref{sec:game}). After a run completion, the user enters three minutes of free play to reestablish baseline physiology. Subsequently, the system varies run difficulty (i.e., low to high or high to low) and begins the cycle again. The user completes four separate runs (two high and two low difficulty) during testing.

\tikzfigcol{raw}{Raw biosignals, collected across three seperate devices, enable downstream feature extraction. (A-B) ECG and Respiration waveforms are collected from the Zephyr Bioharness. (C-E) EDA, PPG, and Skin Temerature are relayed through the Emaptica E4 streaming server. (F) Pupil position is reported through the HTC Vive Pro Eye. A rolling 30s window is applied to each waveform or positional data to extract and report biofeatures (\autoref{t:features}).\vspace{-1.5em}}{

\tikzstyle{component} = [ text centered, rounded corners,font=\Large, inner sep=2, outer sep=0, fill=Goldenrod, text opacity=1, text=black]
\def\blockdist{1}
\def\hblockdist{.5}
\def\vblockdist{1}

\def\imsz{1.1\columnwidth}
\def\botclip{0.1}
\def\lftclip{0.114}
\def\rgtclip{0.1}
\def\topclip{0.15}

\def\toplft{0}
\def\toptop{-0.01}

\def\botlft{0}
\def\botbot{0.05}

\begin{tikzpicture}[ scale=1, transform shape, post/.style={->,shorten >=1pt,>=stealth',semithick, rounded corners=4pt}, font =\tiny]

    
    \path (0,0) node (ecg) [] {\adjincludegraphics[width=\imsz,Clip={{\lftclip\width} {\botclip\height} {\rgtclip\width} {\topclip\height}}]{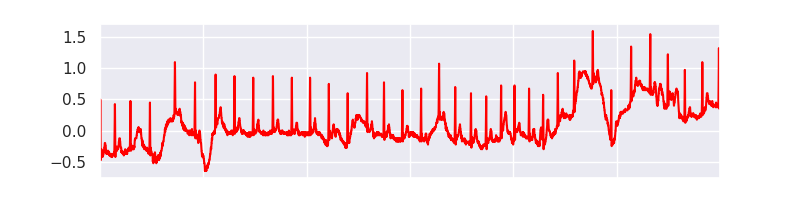}};
    
    \node [align=left, below=.1cm of ecg] (resp)   {\adjincludegraphics[width=\imsz,Clip={{\lftclip\width} {\botclip\height} {\rgtclip\width} {\topclip\height}}]{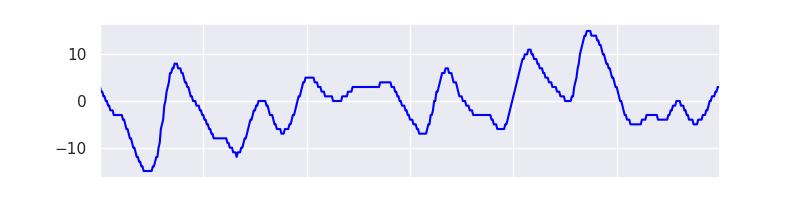}};
    
    \node [below= .1cm of resp] (eda)   {\adjincludegraphics[width=\imsz,Clip={{\lftclip\width} {\botclip\height} {\rgtclip\width} {\topclip\height}}]{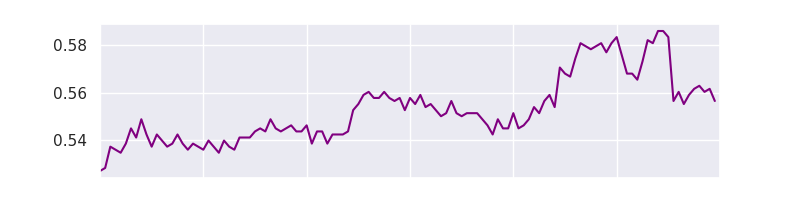}};
   
    \node [align=left, below=.1cm of eda] (ppg)   {\adjincludegraphics[width=\imsz,Clip={{\lftclip\width} {\botclip\height} {\rgtclip\width} {\topclip\height}}]{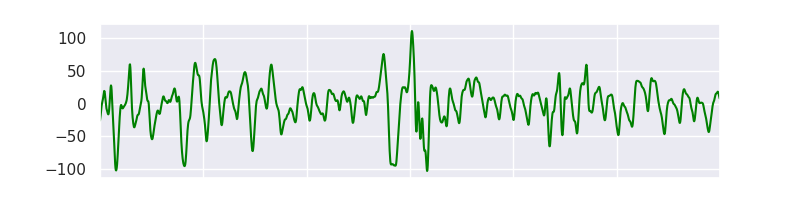}};
    
    \node [align=left, below=.1cm of ppg] (st)   {\adjincludegraphics[width=\imsz,Clip={{\lftclip\width} {0.1\height} {\rgtclip\width} {\topclip\height}}]{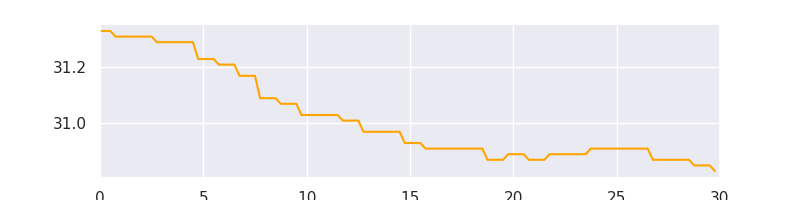}};
    
     \node (pupil) [align=left, below=.2cm of st] {\adjincludegraphics[width=\imsz,Clip={{0.125\width} {0.105\height} {0.1\width} {0.1\height}}]{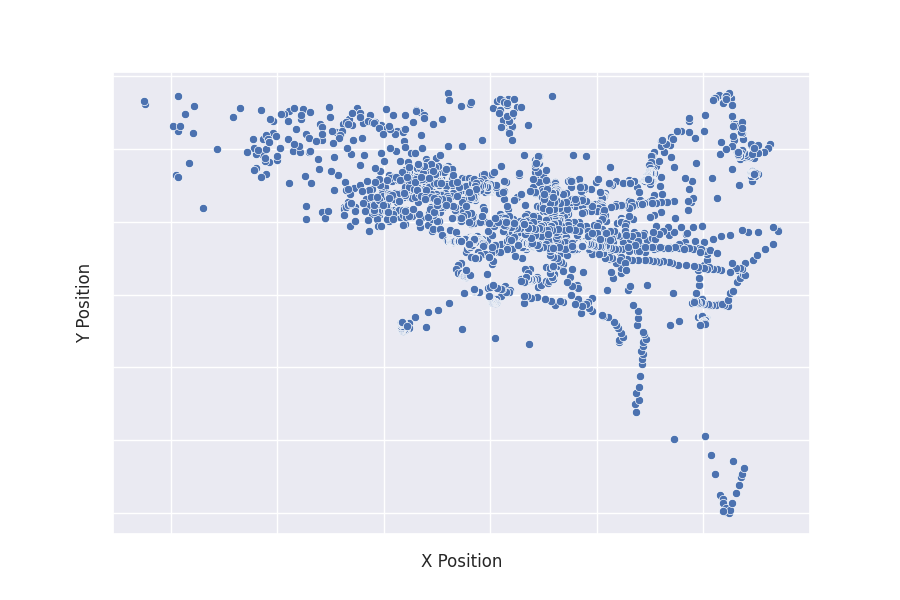}};
   
    \path (ecg.north west) +(\toplft,\toptop) node (topecg) [anchor=north] {1.5};
    \path (ecg.south west) +(\botlft,\botbot) node (botecg) [anchor=south] {-0.5};
    \path ($(topecg) !0.5! (botecg)$) node (ecgunit)[rotate=90] {Amplitude};
   
    \path (eda.north west) +(\toplft,\toptop) node (topeda) [anchor=north] {0.58};
    \path (eda.south west) +(\botlft,\botbot) node (boteda) [anchor=south] {0.55};
    \path ($(topeda) !0.5! (boteda)$) node (edaunit)[rotate=90] {EDA($\mu$S)};
    
    \path (ppg.north west) +(\toplft,\toptop) node (topppg) [anchor=north] {100};
    \path (ppg.south west) +(\botlft,\botbot) node (botppg) [anchor=south] {-100};
    \path ($(topppg) !0.5! (botppg)$) node (ppgunit)[rotate=90] {BVP};
    
    \path (resp.north west) +(\toplft,\toptop) node (topresp) [anchor=north] {10};
    \path (resp.south west) +(\botlft,\botbot) node (botresp) [anchor=south] {-10};
    \path ($(topresp) !0.5! (botresp)$) node (respunit)[rotate=90] {Amplitude};
    
    \path (st.north west) +(\toplft,\toptop) node (topst) [anchor=north] {31.5};
    \path (st.south west) +(\botlft,\botbot) node (botst) [anchor=south] {30.5};
    \path ($(topst) !0.5! (botst)$) node (stunit)[rotate=90] {Temp.($^{\circ}$C)};
    
    \path (pupil.north west) +(\toplft,-.17) node (toppupil) [anchor=north] {45};
    \path (pupil.south west) +(\botlft,\botbot) node (botpupil) [anchor=south] {-45};
    \path ($(toppupil) !0.5! (botpupil)$) node (pupilunit)[rotate=90] {Y Position ($^{\circ}$)};
    
    \path (ecg.north) node (title)[font=\footnotesize] {Electrocardiogram};
    \path (eda.north) node (title)[font=\footnotesize]  {Electrodermal Activity};
    \path (ppg.north) node (title)[font=\footnotesize]  {Photoplethysmography};
    \path (resp.north) node (title)[font=\footnotesize]  {Respiration};
    \path (st.north) node (title)[font=\footnotesize]  {Skin Temperature};
    \path (pupil.north)+(0,-.1) node (title)[font=\footnotesize]  {Pupil Position};
    
    \draw[color=white] ($(st.south west)+(.2,0.14)$)  -- ($(st.south east)+(-.1,0.14)$);
    \path ($(st.south west)+(.25,0.2)$) node (start) [anchor=north] {0};
    \path ($(st.south east)+(-.25,0.19)$) node (end) [anchor=north] {30};
    \path ($(start) !0.5! (end)$) node (time)[] {Time(s)};
    
    \draw[color=white] ($(pupil.south west)+(.2,0.14)$)  -- ($(pupil.south east)+(-.1,0.14)$);
    \path ($(pupil.south west)+(.35,0.2)$) node (start) [anchor=north] {-60};
    \path ($(pupil.south east)+(-.25,0.19)$) node (end) [anchor=north] {60};
    \path ($(start) !0.5! (end)$) node (xpos)[] {X Position ($^{\circ}$)};
    
    \path (topecg.north)+(0,0.05) node (title)[font=\scriptsize] {(A)};
    \path (topresp.north)+(0,0.05) node (title)[font=\scriptsize] {(B)};
    \path (topeda.north)+(0,0.05) node (title)[font=\scriptsize] {(C)};
    \path (topppg.north)+(0,0.05) node (title)[font=\scriptsize] {(D)};
    \path (topst.north)+(0,0.05) node (title)[font=\scriptsize] {(E)};
    \path (toppupil.north)+(0,0.05) node (title)[font=\scriptsize] {(F)};

\end{tikzpicture}
}

\section{System Output}

\ac{ROS} time syncs all collected data across multiple sensors and operating systems (i.e., Linux and Windows). After an individual study completion, a single \textit{rosbag} carries all data for a single run. The \ac{ROS} architecture enables isolation of each data stream to separate \textit{topics}, with each topic handling different measurement types (e.g., \ac{ECG} topic, rover topic, \ac{TLX} survey topic). The system collects in-game objective performance measures, between run \ac{TLX} surveys, and multiple biosignal (e.g., \ac{EDA}, Pupillometry) features.

\subsection{Objective Measures}

In-simulation measures indicate user performance. For the rover, the system reports rover pose (position, rotation), twist (velocity, angular velocity), fuel, and motor temperature at 10 Hz. Boolean flags mark radio requests and responses. Twelve total states represent antenna accuracy. We also collect current CO$_2$ and O$_2$ (at 10 Hz), the current trial leg and difficulty (see \autoref{sec:trial}), and elapsed time of play.

\subsection{Subjective Measures}

A \ac{TLX} survey appears after the user has reached the objective and dropped their marker. The survey asks questions aimed at subjectively evaluating user performance (e.g., How mentally demanding was the task?), in six categories: mental demand, physical demand, temporal demand, performance, effort, and frustration.

\begin{table}[!t] 
\caption{List of biosignals and the corresponding available derived features produced in the \ac{VR} pipeline.}
\label{t:features}
\resizebox{\columnwidth}{!}{
    \begin{tabular}{l|cccccc}
    \hline
    \rowcolor[HTML]{C0C0C0}
    \textbf{Biosignal} & \multicolumn{6}{c}{\textbf{Features}} \\ \hline
    \multirow{7}{*}{\ac{ECG}}        & \multicolumn{2}{c}{\underline{\textit{Time Domain}}} & \multicolumn{2}{c}{\underline{\textit{Non-Linear}}} &\multicolumn{2}{c}{\underline{\textit{Frequency Domain}}} \\
                    & \multicolumn{2}{c}{HRV (RMSSD)} & \multicolumn{2}{c}{PNN10} & \multicolumn{2}{c}{VLF Power} \\
                    & \multicolumn{2}{c}{SDSD}  & \multicolumn{2}{c}{PNN25} & \multicolumn{2}{c}{LF Power} \\
                    & \multicolumn{2}{c}{R2R (RRMean)} & \multicolumn{2}{c}{PNN50} & \multicolumn{2}{c}{HF Power} \\
                    & \multicolumn{2}{c}{RRStD} & \multicolumn{2}{c}{SD1, SD2} & \multicolumn{2}{c}{Total Power} \\
                    & \multicolumn{2}{c}{Min, Max} & \multicolumn{2}{c}{SD1/2}  \\
                    & \multicolumn{2}{c}{HRV Tri Index} & \multicolumn{2}{c}{SDell} \\ \hline
                    
    \multirow{3}{*}{Respiration}     & \multicolumn{3}{c}{\underline{\textit{Time Domain}}} & \multicolumn{3}{c}{\underline{\textit{Frequency Domain}}} \\
                    & \multicolumn{3}{c}{Respiration Rate} & \multicolumn{3}{c}{LF Power, HF Power} \\
                    & \multicolumn{3}{c}{} & \multicolumn{3}{c}{Power Ratio} \\ \hline
                    
    \multirow{5}{*}{\ac{EDA}}        & \multicolumn{2}{c}{\underline{\textit{Phasic}}} & \multicolumn{2}{c}{\underline{\textit{Peak (Phasic)}}} & \multicolumn{2}{c}{\underline{\textit{Tonic}}} \\
                                      & \multicolumn{2}{c}{Mean} &  \multicolumn{2}{c}{Max, Min} &  \multicolumn{2}{c}{Mean} \\
                                      & \multicolumn{2}{c}{StD} &   \multicolumn{2}{c}{Mean, Quantity} &           \multicolumn{2}{c}{StD} \\
                                      & \multicolumn{2}{c}{Range} & \multicolumn{2}{c}{Mean Duration} &          \multicolumn{2}{c}{Range} \\
                                      & \multicolumn{2}{c}{AUC} &   \multicolumn{2}{c}{Mean Slope} &      \multicolumn{2}{c}{AUC} \\ \hline
                                     
    \multirow{2}{*}{\ac{ST}}         & \multicolumn{6}{c}{\underline{\textit{Time Domain}}} \\ 
                                     & \multicolumn{6}{c}{Mean, Median, StD, Max, Min} \\ \hline
    
    \multirow{3}{*}{\ac{PPG}}        &  \multicolumn{6}{c}{\underline{\textit{Time Domain}}} \\
                                        &  \multicolumn{6}{c}{Digital PA, Reflection index} \\
                                        &  \multicolumn{6}{c}{R2R, AUC, sVRI, IPA, PRV} \\ \hline
    \multirow{5}{*}{Pupillometry}   & \multicolumn{6}{c}{\underline{\textit{Time Domain}}} \\ 
                                    & \multicolumn{6}{c}{Pupil Diameter, Smooth Pursuit Frequency} \\
                                    & \multicolumn{6}{c}{Saccade Frequency, Mean Saccade Amplitude} \\
                                    & \multicolumn{6}{c}{Fixation Frequency, Fixation Duration} \\
                                    & \multicolumn{6}{c}{Post Saccadic Oscillations (PSO) Frequency} \\ \hline
    
    \end{tabular} 
}
\end{table}


\subsection{Biofeatures}
\label{sec:biofeatures}
\subsubsection{Raw Biosignals}

Raw biosignals are collected at resolutions specified by the manufacturer. Using the vendor supplied Windows streaming server, the Empatica E4 collects and relays raw \ac{PPG} at 64 Hz as well as \ac{EDA} and \ac{ST} at 4 Hz. The Zephyr Bioharness directly relays a raw respiration waveform at 1.008 Hz and \ac{ECG} waveform at 252 Hz. Pupillometry, provided through the HTC Vive Pro Eye linked to Unity, reports 2-D eye position and pupil diameter at 120Hz.

\subsubsection{Derived Biosignal Features}

Using third-party packages \cite{bizzego2019pyphysio, pekkanen2017new}, our system derives biosignal features with a 30s windowing interval \cite{heard2019multi, tervonen2021ultra}. Biofeatures are derived in both time and frequency domains, ranging from traditional features (e.g., heart rate) to newer approaches (e.g., smooth pursuits). A full list of derived biosignal features can be found in \autoref{t:features}.

\section{Discussion}

 \ac{MW} models are critical for effective \ac{HR} teaming and immersive \ac{VR} experiences, but available processes rarely include online capabilities or relevant settings. This work developed a \ac{VR} pipeline capable of time synchronizing numerous objective and subjective performance data with passive biosignals, which can be utilized in \ac{MW} modeling on or offline (\autoref{f:block}). \ac{VR} task design compels users to focus on the four pillars of \ac{MATB-II} (i.e., tracking, \ac{RM}, \ac{SM}, and communication) while navigating to an objective (\autoref{f:hud}). \ac{VR} enables situational data collection, by emulating high-risk extraplanetary exploration. Raw biosignal data (\autoref{f:raw}) empowers our pipeline to extract a plethora of relevant biofeatures (\autoref{t:features}), close to real time, providing a relevant test bed for open and closed loop \ac{MW} model exploration.

The immersive capability of \ac{VR}, coupled with synchronized biofeatures, enable a deep exploration of \ac{MW} modeling. Immersive environments ensure realistic data collection by causing more realistic mental and physical responses \cite{choi2014effects}. A multitude of biofeatures relate to \ac{MW} (see \autoref{sec:previous_biosignals}), yet few studies directly compare information gain across modalities \cite{vanneste2020towards}. Direct comparisons could inform systems designed for resource constrained environments. \ac{ML} model approaches are end goal dependent (e.g., training evaluation or \ac{HRI}). Online methods, such as \ac{DL} or \ac{RL} models, may improve \ac{HR} teaming efficacy and safety \cite{kim2017intrinsic,le2018efficient}. Offline modeling, such as traditional supervised learning models, may be more appropriate for testing individual skill sets \cite{moustafa2017assessment}. The flexibility of our pipeline design with \ac{ROS} integration allows this approach to be ported to a plethora of \ac{VR} environments, or even into physical robotic systems. 

We learned many lessons and discovered implicit design limitations while executing our system design. Many technical issues stemmed from communicating and synchronizing across two different operating systems (i.e., Windows and Linux). The latest iteration of \ac{ROS}, ROS2, is available for Windows. However, keeping our system in Linux enables downstream use in real-world robotic systems, the majority of which operate on Linux. The Empatica streaming server instability and hardware limitations (e.g., dropping or not discovering devices and specific Bluetooth module requirements) was an unforeseen complication, which could have been remedied with an open-source option; we are currently developing such an option for community benefit. Poor error messaging made debugging the \ac{ROS} to Unity connection difficult (e.g., mistyped Unity components break the connection but do not indicate where the error occurred). Incompatibilities between Unity VR and Steam VR hindered full headset integration, prompting a switch to Unity XR. We tested many iterations in order to balance the \ac{MATB-II} style features to ensure playability and goal completion while being able to modulate difficulty. Our \ac{oop} approach for feature extraction allows easy third-party package changes, ensuring easy future modifications. Other biosensors could be added to the suite (e.g., \ac{EEG}, \ac{EMG}), but may be impractical for \ac{VR} or result in large \textit{rosbag} or \textit{rosmsg} files that can slow down system processing. The 30s windowing time temporally limits \ac{MW} predictions. Shortening the window time without significant information loss may be possible \cite{tervonen2021ultra}.

Simulation design leaves the door open for a variety of future work. An independent samples t-test would validate simulation difficulty against \ac{MW} (i.e., over or under-loaded). We have begun human subject data collection to this end ($\alpha$=0.05, $\beta$=0.8 $\vert$ \textit{n}$\approx$35). With a subject database, we can explore a variety of \ac{MW} \ac{ML} models. We intend to evaluate traditional multivariate linear and simple supervised learning approaches, followed by \ac{DL} and \ac{RL} methods. To utilize the \ac{MW} models online, we are developing an \ac{AI} agent equipped with the \ac{MW} model and capable of online simulation control variation.

\section{Conclusion}

\ac{MW}s could be useful in a wide variety of systems, from \ac{HRI} and teaming to operator training and testing. Such systems would utilize \ac{ML}-based \ac{MW} models that ideally function on passive biosignals produced in realistic environments. \ac{VR}s immersive capabilities facilitate reproducible simulations of complex scenarios, fulfilling this critical need. The developed simulation, based on \ac{CLT} and \ac{MATB-II}, aims to achieve this goal in conjunction with a robust sweep of biosignals and features. The online time synced design not only allows for offline and open loop analysis, but enables real-time \ac{MW} assessment, which stands to improve a variety of \ac{MW} models and form the basis for closed-loop control systems reactive to an operator's \ac{MW}.

\bibliographystyle{abbrv-doi}

\bibliography{main}

\begin{thebibliography}{10}

\bibitem{abdelrahman2017cognitive}
Y.~Abdelrahman, E.~Velloso, T.~Dingler, A.~Schmidt, and F.~Vetere.
\newblock Cognitive heat: exploring the usage of thermal imaging to
  unobtrusively estimate cognitive load.
\newblock {\em Proceedings of the ACM on Interactive, Mobile, Wearable and
  Ubiquitous Technologies}, 1(3):1--20, 2017.

\bibitem{abdurrahman2021effects}
U.~A. Abdurrahman, S.-C. Yeh, Y.~Wong, and L.~Wei.
\newblock Effects of neuro-cognitive load on learning transfer using a virtual
  reality-based driving system.
\newblock {\em Big Data and Cognitive Computing}, 5(4):54, 2021.

\bibitem{albuquerque2020wauc}
I.~Albuquerque, A.~Tiwari, M.~Parent, R.~Cassani, J.-F. Gagnon, D.~Lafond,
  S.~Tremblay, and T.~H. Falk.
\newblock Wauc: a multi-modal database for mental workload assessment under
  physical activity.
\newblock {\em Frontiers in Neuroscience}, 14, 2020.

\bibitem{andrievskaia2020neural}
P.~Andrievskaia, K.~Van~Benthem, and C.~M. Herdman.
\newblock Neural correlates of mental workload in virtual flight simulation.
\newblock In {\em International Conference on Human-Computer Interaction}, pp.
  521--528. Springer, 2020.

\bibitem{backs2000application}
R.~W. Backs.
\newblock Application of psychophysiological models to mental workload.
\newblock In {\em Proceedings of the Human Factors and Ergonomics Society
  Annual Meeting}, vol.~44, pp. 3--464. SAGE Publications Sage CA: Los Angeles,
  CA, 2000.

\bibitem{bailenson2008effect}
J.~Bailenson, K.~Patel, A.~Nielsen, R.~Bajscy, S.-H. Jung, and G.~Kurillo.
\newblock The effect of interactivity on learning physical actions in virtual
  reality.
\newblock {\em Media Psychology}, 11(3):354--376, 2008.

\bibitem{barre2019virtual}
J.~Barr{\'e}, D.~Michelet, J.~Truchot, E.~Jolivet, T.~Recanzone, S.~Stiti,
  A.~Tesni{\`e}re, and G.~Pourcher.
\newblock Virtual reality single-port sleeve gastrectomy training decreases
  physical and mental workload in novice surgeons: an exploratory study.
\newblock {\em Obesity surgery}, 29(4):1309--1316, 2019.

\bibitem{bizzego2019pyphysio}
A.~Bizzego, A.~Battisti, G.~Gabrieli, G.~Esposito, and C.~Furlanello.
\newblock pyphysio: A physiological signal processing library for data science
  approaches in physiology.
\newblock {\em SoftwareX}, 10:100287, 2019.

\bibitem{braithwaite2013guide}
J.~J. Braithwaite, D.~G. Watson, R.~Jones, and M.~Rowe.
\newblock A guide for analysing electrodermal activity ({EDA}) \& skin
  conductance responses ({SCRs}) for psychological experiments.
\newblock {\em Psychophysiology}, 49(1):1017--1034, 2013.

\bibitem{chao2017effects}
C.-J. Chao, S.-Y. Wu, Y.-J. Yau, W.-Y. Feng, and F.-Y. Tseng.
\newblock Effects of three-dimensional virtual reality and traditional training
  methods on mental workload and training performance.
\newblock {\em Human Factors and Ergonomics in Manufacturing \& Service
  Industries}, 27(4):187--196, 2017.

\bibitem{charles2019measuring}
R.~L. Charles and J.~Nixon.
\newblock Measuring mental workload using physiological measures: A systematic
  review.
\newblock {\em Applied ergonomics}, 74:221--232, 2019.

\bibitem{choi2014effects}
H.-H. Choi, J.~J. Van~Merri{\"e}nboer, and F.~Paas.
\newblock Effects of the physical environment on cognitive load and learning:
  Towards a new model of cognitive load.
\newblock {\em Educational Psychology Review}, 26(2):225--244, 2014.

\bibitem{daviaux2019feedback}
Y.~Daviaux, C.~Bey, L.~Arsac, O.~Morellec, and S.~Lini.
\newblock Feedback on the use of {MATB-II} task for modeling of cognitive
  control levels through psycho-physiological biosignals.
\newblock In {\em 20th International Symposium on Aviation Psychology}, p. 205,
  2019.

\bibitem{galy2012relationship}
E.~Galy, M.~Cariou, and C.~M{\'e}lan.
\newblock What is the relationship between mental workload factors and
  cognitive load types?
\newblock {\em International journal of psychophysiology}, 83(3):269--275,
  2012.

\bibitem{grassmann2016respiratory}
M.~Grassmann, E.~Vlemincx, A.~Von~Leupoldt, J.~M. Mittelst{\"a}dt, and
  O.~Van~den Bergh.
\newblock Respiratory changes in response to cognitive load: a systematic
  review.
\newblock {\em Neural plasticity}, 2016.

\bibitem{haapalainen2010psycho}
E.~Haapalainen, S.~Kim, J.~F. Forlizzi, and A.~K. Dey.
\newblock Psycho-physiological measures for assessing cognitive load.
\newblock In {\em Proceedings of the 12th ACM international conference on
  Ubiquitous computing}, pp. 301--310, 2010.

\bibitem{hart2006nasa}
S.~G. Hart.
\newblock {NASA}-task load index ({NASA-TLX}); 20 years later.
\newblock In {\em Proceedings of the human factors and ergonomics society
  annual meeting}, vol.~50, pp. 904--908. Sage publications Sage CA: Los
  Angeles, CA, 2006.

\bibitem{heard2019multi}
J.~Heard and J.~A. Adams.
\newblock Multi-dimensional human workload assessment for supervisory
  human--machine teams.
\newblock {\em Journal of Cognitive Engineering and Decision Making},
  13(3):146--170, 2019.

\bibitem{heard2017human}
J.~Heard, C.~E. Harriott, and J.~A. Adams.
\newblock A human workload assessment algorithm for collaborative human-machine
  teams.
\newblock In {\em 2017 26th IEEE International Symposium on Robot and Human
  Interactive Communication ({RO-MAN})}, pp. 366--371. IEEE, 2017.

\bibitem{holden2013evidence}
K.~Holden, N.~Ezer, and G.~Vos.
\newblock Evidence report: risk of inadequate human-computer interaction.
\newblock 2013.

\bibitem{imhoff2016driving}
S.~Imhoff, M.~Lavalli{\`e}re, N.~Teasdale, and P.~Fait.
\newblock Driving assessment and rehabilitation using a driving simulator in
  individuals with traumatic brain injury: A scoping review.
\newblock {\em NeuroRehabilitation}, 39(2):239--251, 2016.

\bibitem{johnson2014coactive}
M.~Johnson, J.~M. Bradshaw, P.~J. Feltovich, C.~M. Jonker, M.~B. Van~Riemsdijk,
  and M.~Sierhuis.
\newblock Coactive design: Designing support for interdependence in joint
  activity.
\newblock {\em Journal of Human-Robot Interaction, 3 (1), 2014}, 2014.

\bibitem{kakkos2019mental}
I.~Kakkos, G.~N. Dimitrakopoulos, L.~Gao, Y.~Zhang, P.~Qi, G.~K. Matsopoulos,
  N.~Thakor, A.~Bezerianos, and Y.~Sun.
\newblock Mental workload drives different reorganizations of functional
  cortical connectivity between {2D} and {3D} simulated flight experiments.
\newblock {\em IEEE Transactions on Neural Systems and Rehabilitation
  Engineering}, 27(9):1704--1713, 2019.

\bibitem{kaufeld2019level}
M.~Kaufeld and P.~Nickel.
\newblock Level of robot autonomy and information aids in human-robot
  interaction affect human mental workload--an investigation in virtual
  reality.
\newblock In {\em International Conference on Human-Computer Interaction}, pp.
  278--291. Springer, 2019.

\bibitem{kim2017intrinsic}
S.~K. Kim, E.~A. Kirchner, A.~Stefes, and F.~Kirchner.
\newblock Intrinsic interactive reinforcement learning--using error-related
  potentials for real world human-robot interaction.
\newblock {\em Scientific reports}, 7(1):1--16, 2017.

\bibitem{koelstra2011deap}
S.~Koelstra, C.~Muhl, M.~Soleymani, J.-S. Lee, A.~Yazdani, T.~Ebrahimi, T.~Pun,
  A.~Nijholt, and I.~Patras.
\newblock {DEAP}: A database for emotion analysis using physiological signals.
\newblock {\em IEEE transactions on affective computing}, 3(1):18--31, 2011.

\bibitem{krejtz2018eye}
K.~Krejtz, A.~T. Duchowski, A.~Niedzielska, C.~Biele, and I.~Krejtz.
\newblock Eye tracking cognitive load using pupil diameter and microsaccades
  with fixed gaze.
\newblock {\em PloS one}, 13(9):e0203629, 2018.

\bibitem{larmuseau2020multimodal}
C.~Larmuseau, J.~Cornelis, L.~Lancieri, P.~Desmet, and F.~Depaepe.
\newblock Multimodal learning analytics to investigate cognitive load during
  online problem solving.
\newblock {\em British Journal of Educational Technology}, 51(5):1548--1562,
  2020.

\bibitem{le2018efficient}
T.~D. Le, D.~T. Huynh, and H.~V. Pham.
\newblock Efficient human-robot interaction using deep learning with mask
  {R-CNN}: detection, recognition, tracking, and segmentation.
\newblock In {\em 2018 15th International Conference on Control, Automation,
  Robotics and Vision (ICARCV)}, pp. 162--167. IEEE, 2018.

\bibitem{luong2021survey}
T.~Luong, A.~Lecuyer, N.~Martin, and F.~Argelaguet.
\newblock A survey on affective and cognitive {VR}.
\newblock {\em IEEE Transactions on Visualization and Computer Graphics}, 2021.

\bibitem{ma2018human}
L.~M. Ma, T.~Fong, M.~J. Micire, Y.~K. Kim, and K.~Feigh.
\newblock Human-robot teaming: Concepts and components for design.
\newblock In {\em Field and service robotics}, pp. 649--663. Springer, 2018.

\bibitem{marquart2015review}
G.~Marquart, C.~Cabrall, and J.~de~Winter.
\newblock Review of eye-related measures of drivers’ mental workload.
\newblock {\em Procedia Manufacturing}, 3:2854--2861, 2015.

\bibitem{moustafa2017assessment}
K.~Moustafa, S.~Luz, and L.~Longo.
\newblock Assessment of mental workload: a comparison of machine learning
  methods and subjective assessment techniques.
\newblock In {\em International symposium on human mental workload: Models and
  applications}, pp. 30--50. Springer, 2017.

\bibitem{paas2003cognitive}
F.~Paas, J.~E. Tuovinen, H.~Tabbers, and P.~W. Van~Gerven.
\newblock Cognitive load measurement as a means to advance cognitive load
  theory.
\newblock {\em Educational psychologist}, 38(1):63--71, 2003.

\bibitem{paas1994instructional}
F.~G. Paas and J.~J. Van~Merri{\"e}nboer.
\newblock Instructional control of cognitive load in the training of complex
  cognitive tasks.
\newblock {\em Educational psychology review}, 6(4):351--371, 1994.

\bibitem{pekkanen2017new}
J.~Pekkanen and O.~Lappi.
\newblock A new and general approach to signal denoising and eye movement
  classification based on segmented linear regression.
\newblock {\em Scientific reports}, 7(1):1--13, 2017.

\bibitem{pouliquen2016remote}
L.~Pouliquen-Lardy, I.~Milleville-Pennel, F.~Guillaume, and F.~Mars.
\newblock Remote collaboration in virtual reality: asymmetrical effects of task
  distribution on spatial processing and mental workload.
\newblock {\em Virtual Reality}, 20(4):213--220, 2016.

\bibitem{rieger2014heart}
A.~Rieger, R.~Stoll, S.~Kreuzfeld, K.~Behrens, and M.~Weippert.
\newblock Heart rate and heart rate variability as indirect markers of
  surgeons’ intraoperative stress.
\newblock {\em International archives of occupational and environmental
  health}, 87(2):165--174, 2014.

\bibitem{rubio2004evaluation}
S.~Rubio, E.~D{\'\i}az, J.~Mart{\'\i}n, and J.~M. Puente.
\newblock Evaluation of subjective mental workload: A comparison of {SWAT},
  {NASA-TLX}, and workload profile methods.
\newblock {\em Applied psychology}, 53(1):61--86, 2004.

\bibitem{santiago2011multi}
Y.~Santiago-Espada, R.~R. Myer, K.~A. Latorella, and J.~R. Comstock~Jr.
\newblock The {M}ulti-{A}ttribute {T}ask {B}attery {II} ({MATB-II}) software
  for human performance and workload research: A user's guide.
\newblock 2011.

\bibitem{schwalm2008pupillometry}
M.~Schwalm, A.~Keinath, and H.~D. Zimmer.
\newblock Pupillometry as a method for measuring mental workload within a
  simulated driving task.
\newblock {\em Human Factors for assistance and automation}, 1986:1--13, 2008.

\bibitem{sweller1998cognitive}
J.~Sweller, J.~J. Van~Merrienboer, and F.~G. Paas.
\newblock Cognitive architecture and instructional design.
\newblock {\em Educational psychology review}, 10(3):251--296, 1998.

\bibitem{tervonen2021ultra}
J.~Tervonen, K.~Pettersson, and J.~M{\"a}ntyj{\"a}rvi.
\newblock Ultra-short window length and feature importance analysis for
  cognitive load detection from wearable sensors.
\newblock {\em Electronics}, 10(5):613, 2021.

\bibitem{tichon2006training}
J.~Tichon.
\newblock Training cognitive skills in virtual reality: Measuring performance.
\newblock {\em CyberPsychology \& Behavior}, 10(2):286--289, 2006.

\bibitem{vanneste2020towards}
P.~Vanneste, A.~Raes, J.~Morton, K.~Bombeke, B.~B. Van~Acker, C.~Larmuseau,
  F.~Depaepe, and W.~Van~den Noortgate.
\newblock Towards measuring cognitive load through multimodal physiological
  data.
\newblock {\em Cognition, Technology \& Work}, pp. 1--19, 2020.

\bibitem{wang2019data}
C.~Wang and J.~Guo.
\newblock A data-driven framework for learners’ cognitive load detection
  using {ECG-PPG} physiological feature fusion and {XGBoost} classification.
\newblock {\em Procedia computer science}, 147:338--348, 2019.

\bibitem{zepf2020driver}
S.~Zepf, J.~Hernandez, A.~Schmitt, W.~Minker, and R.~W. Picard.
\newblock Driver emotion recognition for intelligent vehicles: A survey.
\newblock {\em ACM Computing Surveys (CSUR)}, 53(3):1--30, 2020.

\bibitem{zhang2018evaluating}
X.~Zhang, Y.~Lyu, X.~Hu, Z.~Hu, Y.~Shi, and H.~Yin.
\newblock Evaluating photoplethysmogram as a real-time cognitive load
  assessment during game playing.
\newblock {\em International Journal of Human--Computer Interaction},
  34(8):695--706, 2018.

\bibitem{zheng2012workload}
B.~Zheng, X.~Jiang, G.~Tien, A.~Meneghetti, O.~N.~M. Panton, and M.~S. Atkins.
\newblock Workload assessment of surgeons: correlation between {NASA-TLX} and
  blinks.
\newblock {\em Surgical endoscopy}, 26(10):2746--2750, 2012.

\end{thebibliography}

\end{document}